\documentclass{article}
\usepackage{amsmath}
\usepackage{graphicx}

\input{tcilatex}
\begin{document}

\title{Analysis and estimation of the threshold for a microwave
\textquotedblleft pellicle mirror\textquotedblright\ parametric oscillator,
via energy conservation}
\date{November 14, 2012}
\author{Raymond Y. Chiao\\Emeritus Professor\\University of California at Merced\\P.O. Box 2039\\Merced,CA 95344\\rchiao@ucmerced.edu}
\maketitle

\begin{figure}[tbp]
\includegraphics[width=5in]{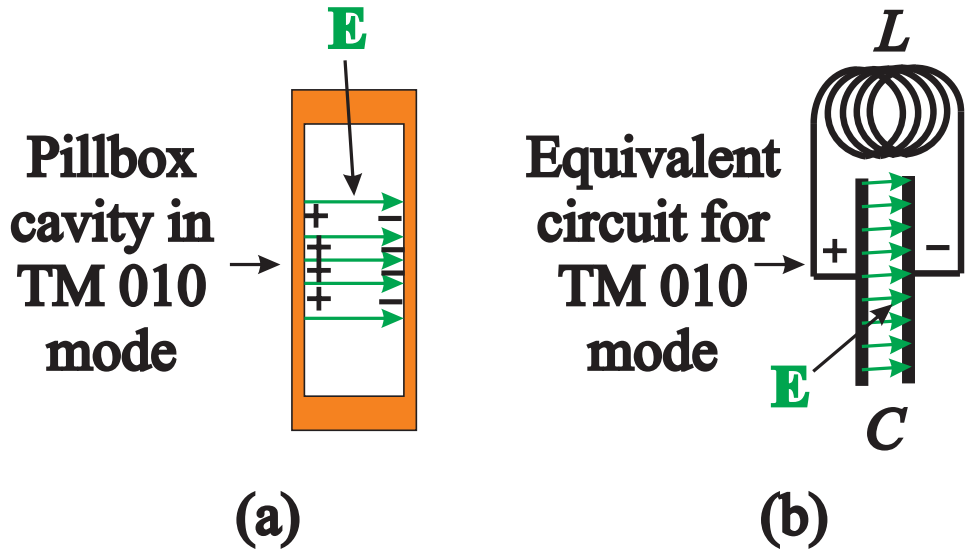}
\caption{(a) The\ microwave TM 010 mode of
a cylindrical \textquotedblleft pillbox cavity\textquotedblright\ pictured
at the instant when the electric field lines $\mathbf{E}$ reach a maximum.
(b) The equivalent $LC$ circuit for this cavity mode, also pictured at the
instant when the electric field lines\ $\mathbf{E}$ reach a maximum inside
the capacitor $C$.}
\end{figure}

{\bf Abstract:} An experiment is proposed to observe the dynamical Casimir effect by means of two tandem, high Q, superconducting microwave cavities, which are separated from each other by only a very thin wall consisting of a flexible superconducting membrane that can be driven into motion by means of resonant ``pump'' microwaves injected into the left cavity. Degenerate ``signal'' and ``idler'' microwave signals can then be generated by the exponential amplification of vacuum fluctuations in the initially empty right cavity, above a certain threshold. The purpose of this paper is calculate the threshold for this novel kind of opto-mechanical parametric oscillation, using energy considerations.\\

In order to be able to numerically estimate the threshold for the
\textquotedblleft pellicle mirror\textquotedblright\ parametric oscillator,
let us start from the $LC$ equivalent circuit for the TM 010 mode of a
cylindrical pillbox cavity\ that is\ illustrated in Figure 1, parts (a) and
(b). Charges of the pillbox cavity in (a) accumulating on opposite sides of
the inner end faces, which produce electric field lines (in green) that run
straight across from the left inner end face to the right inner end face,
can be simulated by charges accumulating on the opposite plates of the
capacitor $C$ in the $LC$ circuit in (b), which produce electric field lines
(in green) that run straight across from the left plate to the right plate.
In other words, we shall model, in a crude first approximation, the \emph{%
nonuniform} electric field inside the TM 010 mode of the pillbox cavity by
means of the \emph{uniform} electric field inside the capacitor $C$ of the $%
LC$ circuit, in order to be able to perform a first, rough estimate of the
threshold.

\begin{figure}[tbp]
\includegraphics[width=5in]{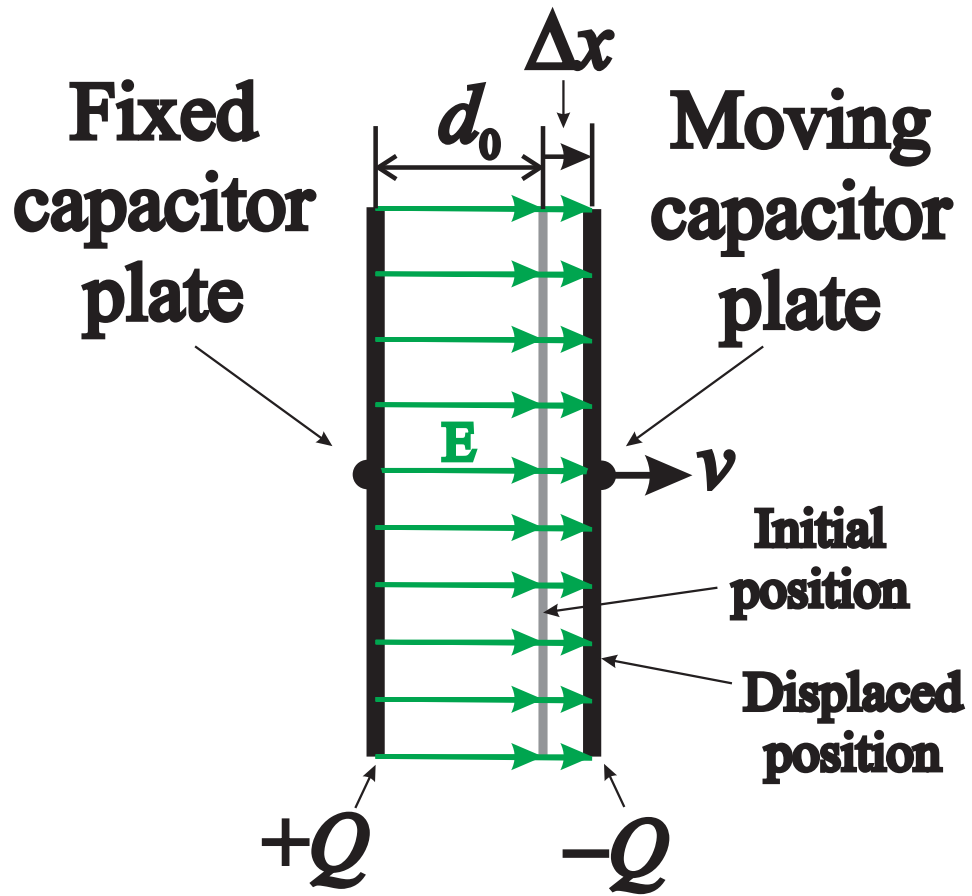}
\caption{Magnified view of the capacitor $C$ in which
the left plate is fixed, but the right plate is displaced towards the right
due to the action of some \textquotedblleft pump.\textquotedblright }
\end{figure}

We shall calculate the threshold using a classical field analysis. Suppose
that the left end face of the pillbox cavity were to be kept fixed, but that
the right end face were to be replaced by a thin, flexible, conducting
\textquotedblleft pellicle mirror\textquotedblright\ that could be driven
into mechanical motion. The analogous situation in the equivalent $LC$
circuit would be for the left plate of the capacitor $C$ to be kept fixed,
but the right plate to be mechanically driven into motion by some external
agency, as illustrated in Figure 2.

For now, let us assume that this motion is so sudden that the charges $+Q$
and $-Q$ on the left and right plates, respectively, of the capacitor $C$,
at the moment of maximum charge accumulation in the $LC$ circuit, cannot
instantaneously change their values during the sudden displacement $\Delta x$%
. This is because the inductor $L$ of the $LC$ circuit in Figure 1, part
(b), forbids any instantaneous change in the current flowing through it at
the instant of maximum charge accumulation pictured in Figure 2.

The work that is done by the external agency\ (the \textquotedblleft
pump\textquotedblright ) on the moving plate of the capacitor that produces
a sudden displacement of an amount $\Delta x$ towards the right, as pictured
in Figure 2, is given by%
\begin{equation}
\Delta W=\left( \frac{1}{2}\varepsilon _{0}E^{2}\right) \cdot A\Delta x
\label{work done on moving capacitor plate}
\end{equation}%
where%
\begin{equation}
u_{E}=\frac{1}{2}\varepsilon _{0}E^{2}
\label{energy density of electric field}
\end{equation}%
is the energy density of the electric field inside the capacitor, and where
the change of the volume $\Delta V$ inside the capacitor due to the sudden
displacement $\Delta x$, is given by%
\begin{equation}
\Delta V=A\Delta x  \label{volume change}
\end{equation}%
where $A$ is the area of the capacitor plate. It is to be stressed that the
sudden displacement $\Delta x$ is caused by some \emph{external} agency.

The mechanical work $\Delta W$ in (\ref{work done on moving capacitor plate}%
) done by this external agency on the fields inside the $LC$ circuit can be
rewritten in the following form%
\begin{equation}
\Delta W=P\Delta V  \label{PdV work done by piston}
\end{equation}%
where the pressure $P$ on the plate of the capacitor is given by%
\begin{equation}
P=\frac{1}{2}\varepsilon _{0}E^{2}  \label{pressure on plate}
\end{equation}%
which is equal to the instantaneous energy density $u_{E}$ given by (\ref%
{energy density of electric field}) \cite{alternative}. Therefore the work
in (\ref{PdV work done by piston}) can be viewed as the \emph{negative} of
the usual thermodynamic \textquotedblleft $PdV$\textquotedblright\ work done 
\emph{by} a closed \textquotedblleft system\textquotedblright\ (here, the
electric fields inside the capacitor) \cite{sign of work}\cite{photon gas} 
\emph{on} the \textquotedblleft environment\textquotedblright\ (here, the
external agency). In this way, one can view the moving \textquotedblleft
pellicle mirror,\textquotedblright\ i.e., the moving end wall of the closed
pillbox cavity, or the moving capacitor plate in the equivalent $LC$
circuit, as if it were a moving piston acting on a closed thermodynamic
system, namely, the vacuum fluctuations inside the cavity, or equivalently,
inside the $LC$ circuit.

The pressure $P=\frac{1}{2}\varepsilon _{0}E^{2}$ in (\ref{pressure on plate}%
) can also be viewed as being one component of the Maxwell stress tensor 
\cite{jackson} 
\begin{equation}
T_{ij}=\varepsilon _{0}\left[ E_{i}E_{j}+c^{2}B_{i}B_{j}-\frac{1}{2}\delta_{i,j}\left( 
\mathbf{E\cdot E}+c^{2}\mathbf{B\cdot B}\right) \right]
\label{Maxwell stress tensor}
\end{equation}%
namely, the component with spatial indices $i=j=3$, i.e., the tensor product
of the \emph{longitudinal} electric field at the surface of the plate with
itself, at the instant when no magnetic fields are present.

Note that there are no electrical charges within the volume $\Delta
V=A\Delta x$ which is being swept out by the motion of the right plate in
Figure 2. Therefore, using Maxwell's first equation%
\begin{equation}
\mathbf{\nabla \cdot E}=\frac{\rho }{\varepsilon _{0}}=0  \label{div E = rho}
\end{equation}%
and using a Gaussian pillbox argument, we conclude that the lines of
electric field $\mathbf{E}$, which originated from the charge $+Q$ on the
left plate, must continue straight across this newly created volume, and
terminate on the charges $-Q$ on the right plate at its new, displaced
position, as indicated in Figure 2. This implies that there must exist a
process of \textquotedblleft creation out of nothing\textquotedblright\ of
new electric field lines of $E$ inside the new volume $\Delta V$, and
therefore that, during this process, there must also exist an accompanying
creation of an extra amount of electromagnetic energy in the $LC$ circuit,%
\begin{equation}
\Delta U_{E}=\left( \frac{1}{2}\varepsilon _{0}E^{2}\right) \Delta V
\label{EM energy created}
\end{equation}%
seemingly out of nothing, due to the sudden displacement of this plate to
the right by the amount $\Delta x$.

However, energy must be conserved. Therefore the meaning of (\ref{work done
on moving capacitor plate}) is it is that the amount of \emph{mechanical} work $\Delta W$ done on the plate must be exactly equal to the amount of \emph{%
electrical} energy $\Delta U_{E}$\ in (\ref{EM energy created}), which is
being created within the $LC$ system by the external \textquotedblleft
pump\textquotedblright\ agency, upon the sudden displacement $\Delta x$ of
the moving plate towards the right.

Now if energy were to be continually supplied to the $LC$ circuit by a
periodic sequence of displacements $\Delta x\left( t\right) $ to the mirror
by some continuous-wave external pumping agency at a correct frequency,
exponential amplification of some seed waveform within the circuit could
result, if a certain synchronization condition is met. Here we shall assume
that the $LC$ circuit is on resonance at the \emph{same} frequency for both
signal and idler waves, i.e., we are only considering a \emph{degenerate}
parametric amplifier. Then the synchronization condition follows from the
fact that in the expression for the electromagnetic energy $\Delta
U_{E}=\left( \frac{1}{2}\varepsilon _{0}E^{2}\right) \Delta V$ inside the
resonator, the factor of $E^{2}$ implies that there would be a second
harmonic term in the pressure $P$ in (\ref{pressure on plate}). Hence if we
choose the origin in time such that the waveform for the electric field
inside the capacitor is given by%
\begin{equation}
E=E_{0}\cos \omega t
\end{equation}%
then it follows that the square of the electric field is given by

\begin{equation}
E^{2}=E_{0}^{2}\cos ^{2}\omega t=E_{0}^{2}\cdot \frac{1}{2}(1+\cos 2\omega t)
\label{trig identity for 2nd harmonic}
\end{equation}%
where $\omega $ is the resonance frequency of the $LC$ circuit, and
therefore that the pressure on the plate in (\ref{pressure on plate})%
\begin{equation}
P=\frac{1}{2}\varepsilon _{0}E^{2}=\frac{1}{4}\varepsilon
_{0}E_{0}^{2}(1+\cos 2\omega t)  \label{pressure with 2nd harmonic term}
\end{equation}%
has a second harmonic term that could then be made synchronous with the
second harmonic pumping term in the mechanical work $\Delta W$ in (\ref{work
done on moving capacitor plate}). We can do so by choosing the periodic
displacement waveform $\Delta x\left( t\right) $ driven by the pump source
to concide in frequency with the second harmonic term in the pressure on the
plate in (\ref{pressure with 2nd harmonic term}).

This motivates us to postulate that the external pump agency constrains the
right plate of the capacitor $C$ in Figure 2 to move so that its velocity
moves at the second harmonic frequency according to the expression%
\begin{equation}
v\left( t\right) =\lim_{\Delta t\rightarrow 0}\frac{\Delta x}{\Delta t}%
=v_{2\omega }\cos 2\omega t  \label{velocity 2nd harmonic form}
\end{equation}%
where $v_{2\omega }$ is a fixed drive velocity amplitude that is determined
by the ``pump.'' Then there arises from the mechanical work in (\ref{work done
on moving capacitor plate}) an expression for the instantaneous mechanical
power, which contains a time-varying factor%
\begin{eqnarray}
\lim_{\Delta t\rightarrow 0}\frac{\Delta W}{\Delta t} &=&\lim_{\Delta
t\rightarrow 0}\left( \frac{1}{2}\varepsilon _{0}E^{2}\left( t\right)
\right) A\frac{\Delta x\left( t\right) }{\Delta t}=\left( \frac{1}{2}%
\varepsilon _{0}E^{2}\left( t\right) \right) Av\left( t\right)  \notag \\
&=&\left( \frac{1}{2}\varepsilon _{0}E_{0}^{2}\right) \cos ^{2}\omega t\cdot
Av_{2\omega }\cos 2\omega t  \notag \\
&=&\left( \frac{1}{4}\varepsilon _{0}E_{0}^{2}\right) (1+\cos 2\omega
t)\cdot Av_{2\omega }\cos 2\omega t  \notag \\
&=&\left( \frac{1}{4}\varepsilon _{0}E_{0}^{2}\right) Av_{2\omega }\left(
\cos 2\omega t+\cos ^{2}2\omega t\right)  \label{instantaneous power}
\end{eqnarray}%
which has a nonzero time average, since the time average of the last term in
(\ref{instantaneous power}) gives%
\begin{equation}
\left\langle \cos ^{2}2\omega t\right\rangle =\frac{1}{2}
\end{equation}%
where the angular brackets denote an average over time.

It therefore follows that the time-averaged power transfer from the external
pump source into the $LC$ circuit is%
\begin{equation}
\left\langle \frac{dW}{dt}\right\rangle =\left( \frac{1}{8}\varepsilon
_{0}E_{0}^{2}\right) Av_{2\omega }  \label{time-average power transfer}
\end{equation}%
Since the time-averaged energy density stored inside the capacitor is%
\begin{equation}
\left\langle u_{E}\right\rangle =\frac{1}{2}\varepsilon
_{0}E_{0}^{2}\left\langle \cos ^{2}\omega t\right\rangle =\frac{1}{4}%
\varepsilon _{0}E_{0}^{2}
\end{equation}%
we can rewrite (\ref{time-average power transfer}) as follows:%
\begin{equation}
\left\langle \frac{dW}{dt}\right\rangle =\frac{1}{2}\left\langle
u_{E}\right\rangle Av_{2\omega }
\label{time-averaged power in terms of second harmonic drive}
\end{equation}%
\qquad

In other words, energy is being continuously transferred from the external
pump source into the $LC$ circuit, with an average rate of transfer given by%
\begin{equation}
\left\langle \frac{dW}{dt}\right\rangle =\left\langle \left( \frac{1}{2}%
\varepsilon _{0}E^{2}\right) A\frac{dx}{dt}\right\rangle =A\left\langle u_{E}%
\frac{dx}{dt}\right\rangle =A\left\langle u_{E}v\right\rangle =\left\langle
F\cdot v\right\rangle  \label{<F.v>}
\end{equation}%
where%
\begin{equation}
v=\frac{dx}{dt}=v_{2\omega }\cos 2\omega t  \label{instantaneous velocity}
\end{equation}%
is the instantaneous velocity of the moving plate fixed by the external
``pump,'' and where $F\left( t\right) =P\left( t\right) A$ is the instantaneous
force on this plate arising from the instantaneous pressure%
\begin{equation}
P\left( t\right) =\frac{1}{2}\varepsilon _{0}E^{2}\left( t\right) =\frac{1}{4%
}\varepsilon _{0}E_{0}^{2}(1+\cos 2\omega t)  \label{time-varying pressure}
\end{equation}%
arising from the electric field inside the capacitor.

In the absence of any dissipation in the $LC$ circuit, it follows from
energy conservation that%
\begin{equation}
\left\langle \frac{dW}{dt}\right\rangle =\frac{d\left\langle
U_{E}\right\rangle }{dt}+\frac{d\left\langle U_{B}\right\rangle }{dt}
\label{energy conservation}
\end{equation}%
Now it can be shown that the time-averaged energy stored in capacitor\ $C$
of the $LC$ circuit is equal to that stored in the inductor $L$, i.e.,%
\begin{equation}
\left\langle U_{E}\right\rangle =\left\langle U_{B}\right\rangle
\label{equipartition}
\end{equation}%
Therefore from (\ref{energy conservation}), (\ref{equipartition}), and (\ref%
{time-averaged power in terms of second harmonic drive}), it follows that%
\begin{equation}
2\frac{d\left\langle U_{E}\right\rangle }{dt}=\left\langle \frac{dW}{dt}%
\right\rangle =\frac{1}{2}\left\langle u_{E}\right\rangle Av_{2\omega
}=+2\kappa \left\langle U_{E}\right\rangle  \label{ODE for U_E}
\end{equation}%
for some constant $\kappa $ to be determined below. The ODE in (\ref{ODE for
U_E}) implies that there will be a continuous energy transfer from the pump
source that will result in an \emph{exponential} growth of the time-averaged
stored energy $\left\langle U_{E}\right\rangle $ stored in the capacitor
(and thus of the total energy in the $LC$ circuit).

To find the proportionality constant $\kappa $, we note that the
relationship between the total energy $U_{E}$\ stored in the capacitor and
the energy \emph{density} $u_{E}$\ stored in the capacitor, is given by%
\begin{equation*}
\left\langle U_{E}\right\rangle =\left\langle u_{E}\right\rangle
Ax_{0}=\left\langle u_{E}\right\rangle V_{0}
\end{equation*}%
where the equilibrium volume $V_{0}$ of the capacitor is given by%
\begin{equation}
V_{0}=Ad_{0}  \label{volume of the cavity}
\end{equation}%
where $d_{0}$ is the equilibrium spacing between the capacitor plates (see
Figure 2). Therefore the ODE for $\left\langle U_{E}\right\rangle $ in (\ref%
{ODE for U_E}) becomes 
\begin{equation}
2Ax_{0}\frac{d\left\langle u_{E}\right\rangle }{dt}=\frac{1}{2}\left\langle
u_{E}\right\rangle Av_{2\omega }=+2\kappa \left\langle u_{E}\right\rangle
Ad_{0}
\end{equation}%
Dividing by $2V_{0}=2Ad_{0}$, we obtain an ODE for $\left\langle
u_{E}\right\rangle $%
\begin{equation}
\frac{d\left\langle u_{E}\right\rangle }{dt}=\frac{1}{4}\left\langle
u_{E}\right\rangle \frac{v_{2\omega }}{d_{0}}=+\kappa \left\langle
u_{E}\right\rangle  \label{ODE for gain without dissipation}
\end{equation}%
Therefore we conclude that the exponential gain coefficient $\kappa $ for
the energy density in (\ref{ODE for gain without dissipation}) is given by%
\begin{equation}
\kappa =\frac{1}{4}\frac{v_{2\omega }}{d_{0}}=\frac{1}{4}\frac{v_{\text{pump}%
}}{d_{0}}  \label{gain = A x velocity}
\end{equation}%
where%
\begin{equation}
v_{2\omega }=v_{\text{pump}}
\label{velocity at second harmonic is pump velocity amplitude}
\end{equation}%
is the amplitude of the second-harmonic component of the instantaneous
velocity (\ref{instantaneous velocity})\ in the mechanical motion driven by
a fixed, \emph{undepleted} pump waveform that is determined by some external
source.

Next, let us introduce a phenomenological loss coefficient $\gamma $ into
the ODE for $\left\langle u_{E}\right\rangle $ in (\ref{ODE for gain without
dissipation}) as follows:%
\begin{equation}
\frac{d\left\langle u_{E}\right\rangle }{dt}=+\kappa \left\langle
u_{E}\right\rangle -\gamma \left\langle u_{E}\right\rangle
\label{ODE with gain and loss}
\end{equation}%
If there is no gain (i.e., if $\kappa =0$), then (\ref{ODE with gain and
loss}) becomes%
\begin{equation}
\frac{d\left\langle u_{E}\right\rangle }{dt}=-\gamma \left\langle
u_{E}\right\rangle =-\frac{1}{\tau }\left\langle u_{E}\right\rangle
\end{equation}%
where the \textquotedblleft cavity ring-down time\textquotedblright\ $\tau $
is given by%
\begin{equation}
\tau =\frac{1}{\gamma }=\frac{Q}{\omega }  \label{tau = 1/gamma}
\end{equation}%
where the quality factor of the cavity $Q$ is defined through%
\begin{equation}
Q=\omega \tau
\end{equation}

The threshold for parametric oscillation, like that in a laser, occurs when
the gain equals the loss, i.e., when%
\begin{equation}
\kappa =\gamma  \label{gain = loss}
\end{equation}%
From (\ref{gain = loss}), (\ref{tau = 1/gamma}), and (\ref{gain = A x
velocity}), we conclude that the threshold condition for the pump to be able
to generate degenerate parametric oscillation, based on the moving
capacitor\ plate model of Figure 2 for the $LC$ equivalent circuit for the
\textquotedblleft pillbox cavity\textquotedblright\ of Figure 1, is given by%
\begin{equation}
\kappa _{\text{threshold}}=\frac{1}{4}\frac{v_{\text{threshold}}}{d_{0}}%
=\gamma =\frac{\omega }{Q}  \label{threshold in terms of Q}
\end{equation}%
where $v_{\text{threshold}}$ is the amplitude of the second-harmonic
component of the velocity of the moving mirror that is being driven by the
second harmonic frequency $2\omega $ of some external pump source, and where $\omega
=\omega _{s}=\omega _{i}$ is the degenerate signal (or idler) frequency of
the $LC$ resonator. Solving for the velocity $v_{\text{threshold}}$ from (%
\ref{threshold in terms of Q}), we find that the minimum velocity amplitude
for the mirror necessary for the onset of degenerate parametric oscillation,
is%
\begin{equation}
v_{\text{threshold}}=\frac{4\omega d_{0}}{Q}
\label{velocity driven by pump at threshold}
\end{equation}

We can then calculate the minimum, time-averaged kinetic energy of the
moving mirror caused by the pump at threshold as follows:%
\begin{equation}
\left\langle K_{\text{threshold}}\right\rangle =\frac{1}{2}m\left\langle v_{%
\text{threshold}}^{2}\right\rangle =\frac{1}{4}m\left( \frac{4\omega d_{0}}{Q%
}\right) ^{2}=\frac{4m\omega ^{2}d_{0}^{2}}{Q^{2}}
\label{kinetic energy at threshold}
\end{equation}%
The total stored energy $U_{\text{threshold}}$ at threshold in the motion of
the mirror is the sum of the threshold kinetic and potential energies, which
are equal to each other for simple harmonic motion, so that%
\begin{equation}
U_{\text{threshold}}=\left\langle K_{\text{threshold}}\right\rangle
+\left\langle V_{\text{threshold}}\right\rangle =2\left\langle K_{\text{%
threshold}}\right\rangle
\end{equation}

Therefore we conclude that the total pump energy required at threshold for
exciting degenerate parametric oscillations of the \textquotedblleft
pellicle mirror\textquotedblright\ is%
\begin{equation}
U_{\text{threshold}}=\frac{8m\omega ^{2}d_{0}^{2}}{Q^{2}}
\label{Degenerate parametric oscillator threshold}
\end{equation}%
where $m$ is the mass of the \textquotedblleft pellicle
mirror\textquotedblright , $\omega $ is the degenerate signal (or idler)
frequency of the generated microwaves in the \textquotedblleft pillbox
cavity\textquotedblright\ mode, and $d_{0}$ is the effective gap of the
capacitance in the equivalent $LC$ circuit for this cavity mode.

Let us compare this expression with the following expression for the
threshold given by Braginsky et al. \cite{Braginsky}\ for the generation of
the parametric oscillations of an elastic mode of a moving mirror of a
Fabry-Perot cavity excited by radiation pressure from a pump laser:%
\begin{equation}
U_{\text{Braginsky's threshold}}=\frac{1}{2}\frac{m\omega _{s}^{2}L^{2}}{%
Q_{i}Q_{s}}  \label{Braginsky's theshold}
\end{equation}%
where $m$ is the mass of the moving mirror, $\omega _{s}$ is the signal
frequency, $L$ is the length of the Fabry-Perot cavity, and $Q_{i}$ and $%
Q_{s}$ are respectively the quality factors of the idler and signal modes of
the parametric oscillator. We see that the expressions for the two
thresholds (\ref{Degenerate parametric oscillator threshold})\ and (\ref%
{Braginsky's theshold}) become identical, if one identifies%
\begin{equation}
d_{0}=\frac{1}{4}L\text{ and }Q_{i}=Q_{s}=Q
\end{equation}%
i.e., if the effective length scale of the capacitor $d_{0}$ in the $LC$
model of the \textquotedblleft pillbox cavity\textquotedblright\ in its TM
010 mode is identified with one quarter of the length $L$ of a Fabry-Perot
resonator in one of its TEM modes, and if the two quality factors$\
Q_{i}=Q_{s}$ become identical, since the signal and idler are being excited
in the same, degenerate mode in the \textquotedblleft pillbox
cavity\textquotedblright\ case. In any case, the two thresholds (\ref%
{Degenerate parametric oscillator threshold}) and (\ref{Braginsky's theshold}%
) are comparable to each other in terms of orders of magnitude, which will
be sufficient for designing experiments.

Going back to the original \textquotedblleft pillbox\textquotedblright\
configuration of Figure 1(a), we estimate that the characteristic length
scale for the \textquotedblleft pillbox cavity\textquotedblright\ is \ 
\begin{equation}
d_{0}\simeq \lambda
\end{equation}%
which would be more appropriate for the TM 012 mode than the TM 010 mode. Now
since%
\begin{equation}
d_{0}\omega \simeq \lambda \omega \simeq 2\pi \lambda f\simeq 2\pi c
\end{equation}%
where $\lambda $ is the wavelength of microwaves with a resonance frequency $%
f=\omega /2\pi $, and $c$ is the speed of light, we conclude that a rough
estimate of the threshold for degenerate parametric oscillation in (\ref%
{velocity driven by pump at threshold}) is given by the expression%
\begin{equation}
\frac{v_{\text{pump}}}{c}\simeq \frac{2\pi }{Q}  \label{v/c at threshold}
\end{equation}%
where $v_{\text{pump}}$ is the velocity of the moving mirror which is being
driven at the second harmonic frequency by the pump.

Let us compare the condition (\ref{v/c at threshold}) with the threshold
condition introduced by Walls and Milburn \cite{Walls and Milburn 2008} in
connection with the production of squeezed states in parametric oscillation
(as cited by Nation et al. \cite{Nation})%
\begin{equation}
\frac{\varepsilon \omega Q}{c}\geq 1
\label{Walls and Millburn's threshold condition}
\end{equation}%
where $\varepsilon $ is the displacement amplitude of the moving mirror
inside a Fabry-Perot resonator (with a fixed mirror and a moving mirror),
whose resonance frequency is $\omega $ and whose quality factor is $Q$. The
velocity amplitude of the sinusoidal motion of the mirror is given by%
\begin{equation}
v_{\text{mirror}}=\varepsilon \omega
\end{equation}%
Therefore it follows that Walls and Millburn's threshold condition (\ref%
{Walls and Millburn's threshold condition}) can be re-expressed in terms of
the mirror's maximum velocity $v_{\text{mirror}}$\ as follows:%
\begin{equation}
\frac{v_{\text{mirror}}}{c}\geq \frac{1}{Q}
\label{v/c for Walls and Milburn}
\end{equation}%
which is indeed consistent with (\ref{v/c at threshold}), when viewed as an
order-of-magnitude estimate of the threshold.

Since for superconducting cavities, it is possible to achieve the extremely
high $Q$'s of the order of \cite{Haroche}%
\begin{equation}
Q\simeq 10^{10}
\end{equation}%
the two threshold conditions (\ref{v/c at threshold}) and (\ref{v/c for
Walls and Milburn}) indicate that it is not necessary for the moving mirror
to be driven into relativistic motion (i.e., with $v_{\text{mirror}}\simeq c$%
) by the pump source in order to achieve the threshold for parametric
oscillation.

\begin{figure}[tbp]
\includegraphics[width=6in]{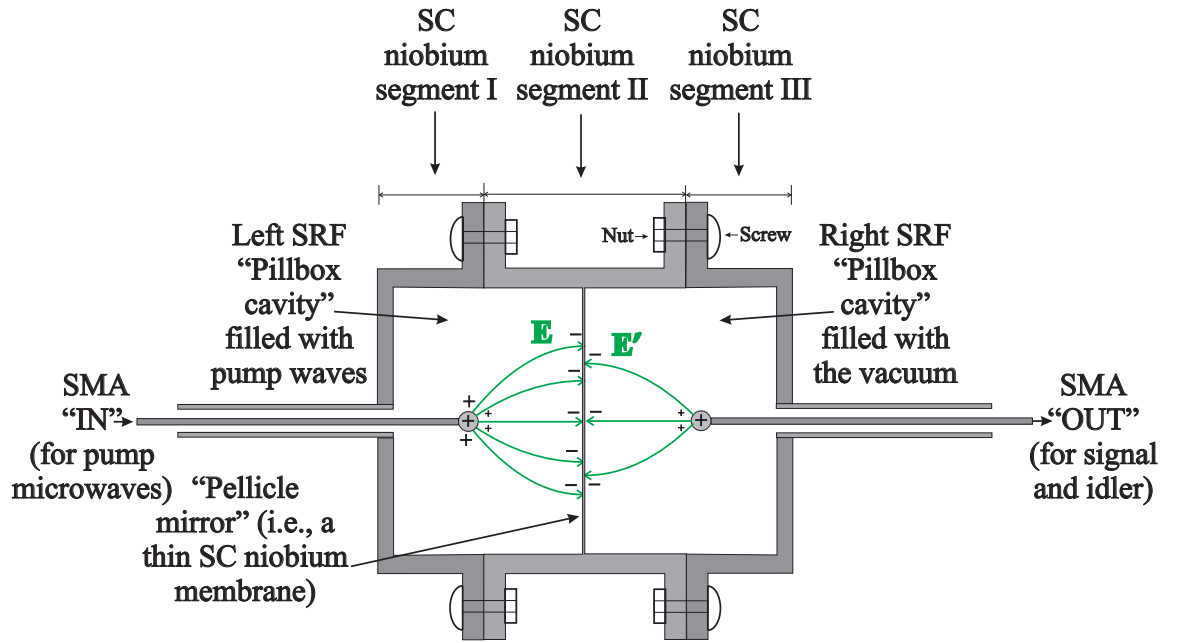}
\caption{Degenerate parametric
oscillator consisting of two superconducting radio frequency (SRF)
\textquotedblleft pillbox cavities\textquotedblright\ bolted together from
three cylindrical segments (I, II, III) of superconducting (SC) niobium. The
central segment (II) has a flexible \textquotedblleft pellicle
mirror\textquotedblright\ incorporated into its midsection as the active
element of the amplifier. The \textquotedblleft pellicle
mirror\textquotedblright\ is a SC niobium membrane that serves as a thin
wall that separates the entire assembly into left and right SC niobium
cavities. The membrane can be charged on its two surfaces by electrostatic
induction. The left cavity is pumped by microwaves injected into the
\textquotedblleft IN\textquotedblright\ port. Degenerate signal and idler
microwaves grow exponentially out of vacuum fluctuations in the right
cavity, and leave through the \textquotedblleft OUT\textquotedblright\ port.}
\end{figure}

One practical way to drive the \textquotedblleft pellicle
mirror\textquotedblright\ into motion is to use a second high-Q microwave
cavity to drive it mechanically via the Maxwell stress tensor on the other
side of this mirror, so as to excite it at the second harmonic of the
degenerate signal-idler cavity frequency (see Figure 3). For example, the
\textquotedblleft pillbox cavity\textquotedblright\ on the left side of the
\textquotedblleft pellicle mirror\textquotedblright\ could be this second
cavity, which could be filled pump microwaves, whereas the \textquotedblleft
pillbox cavity\textquotedblright\ on the right side of the \textquotedblleft
pellicle mirror\textquotedblright , which is initially filled only with the
vacuum, could be filled with degenerate signal and idler microwaves upon
exponential gain above threshold. The left cavity could be excited by means
of an antenna consisting of an extension of the center conductor of the
on-axis \textquotedblleft IN\textquotedblright\ SMA\ cable. This cable could
be charged with a DC bias potential so as to induce electrostatic charges on
the left inner surface of the \textquotedblleft pellicle
mirror\textquotedblright . Similarly, the radiation in the right cavity,
which could be produced out of vacuum fluctuations in the dynamical Casimir
effect, could be extracted by means of an output antenna which is the
extension of the center conductor of the \textquotedblleft
OUT\textquotedblright\ SMA cable sticking out into this cavity. For the design of the nondegenerate parametric oscillator, see \cite{nondegenerate}.

Let us calculate the threshold in terms of the external microwave pump power
that needs to be injected through the \textquotedblleft
IN\textquotedblright\ port of the pump (i.e., left) cavity of Figure 3. To
do so, we note that in steady-state equilibrium, the injected pump power
coming in through the \textquotedblleft IN\textquotedblright\ port at
threshold must be balanced exactly by the lost pump power, i.e.,%
\begin{equation}
\mathcal{P}_{\text{threshold}}=\frac{\left\langle U_{p,\text{threshold}%
}\right\rangle }{\tau _{p}}=\frac{\omega _{p}\left\langle U_{p,\text{%
threshold}}\right\rangle }{Q_{p}}  \label{steady-state pump power}
\end{equation}%
where $\tau _{p}=Q_{p}/\omega _{p}$ is the cavity ring-down time of the pump
cavity.

In Figure 3, the external agency driving the \textquotedblleft pellicle
mirror\textquotedblright\ is the force associated with the Maxwell stress
tensor, which gives rise to a pressure from the pump waveform that is being
injected into the pump (left) cavity. Note that this force is exerted on the 
\emph{opposite} side of \textquotedblleft pellicle mirror\textquotedblright\
from the side that faces the signal (or idler) cavity, and is related to the
acceleration of this \textquotedblleft moving plate\textquotedblright\ in
accordance with Newton's 2nd law%
\begin{equation}
F=PA=\left( \frac{1}{2}\varepsilon _{0}E^{2}\right) A=m\frac{d^{2}x}{dt^{2}}
\label{Newton's 2nd law}
\end{equation}%
where $P=\frac{1}{2}\varepsilon _{0}E^{2}$ is the pressure applied to the
\textquotedblleft plate,\textquotedblright\ $A$ is the area of the
\textquotedblleft plate,\textquotedblright\ $E$ is the total instantaneous
electric field from the pump microwaves being applied at the
\textquotedblleft plate,\textquotedblright\ $m$ is the mass of the
\textquotedblleft plate,\textquotedblright\ and $x$ is the instantaneous
displacement of the \textquotedblleft plate.\textquotedblright\ To a good
approximation, the moving mirror (i.e., the moving plate of the capacitor in
Figure 2) is moving as if it were a \emph{free} mass that is being driven by
the pump microwaves.

First, we shall consider the case in Figure 3 in which there is no external
DC bias injected through the \textquotedblleft IN\textquotedblright\ port
along with the microwave pump field. Let this pump wave have the waveform%
\begin{equation}
E=E_{p}\sin \omega _{p}t  \label{E_p without DC bias}
\end{equation}%
where $E_{p}$ is the electric field amplitude of the pump wave, and the pump
frequency $\omega _{p}$ is at the \emph{first} harmonic (i.e., at the \emph{%
same }frequency as the degenerate signal and idler frequencies $\omega
_{s}=\omega _{i}$). However, note that the left and right cavities form two
separate Faraday cages, so that there is no possibility of any leakage of
pump radiation from the left cavity into in the right cavity, which could be
confused with the signal (or idler) radiation that is generated by the
dynamical Casimir effect.

The square of $E$, which enters into (\ref{Newton's 2nd law}), becomes%
\begin{equation}
E^{2}=E_{p}^{2}\sin ^{2}\omega _{p}t=E_{p}^{2}\left( \frac{1}{2}-\frac{1}{2}%
\sin 2\omega _{p}t\right)  \label{square of E_p without DC bias}
\end{equation}%
which generates the second harmonic needed for pumping the degenerate
parametric oscillations in the right cavity. Looking only the second
harmonic component of Newton's equation of motion (\ref{Newton's 2nd law}),
we see that%
\begin{equation}
F\left( \text{at }2\omega _{p}\right) =\left( -\frac{1}{4}\varepsilon
_{0}E_{p}^{2}\sin 2\omega _{p}t\right) A=m\frac{d^{2}x}{dt^{2}}=-m\left(
2\omega _{p}\right) ^{2}x
\label{2nd harmonic component of Newtons's 2nd law}
\end{equation}%
where the solution for the time-varying displacement is%
\begin{equation}
x=x_{p}\sin 2\omega _{p}t  \label{solution x(t)}
\end{equation}%
where, from (\ref{2nd harmonic component of Newtons's 2nd law}), we find
that the solution for the displacement amplitude is%
\begin{equation}
x_{p}=\frac{1}{16}\frac{\varepsilon _{0}E_{p}^{2}}{m\omega _{p}^{2}}
\label{x_p =...}
\end{equation}%
The solution for the time-varying velocity of the driven \textquotedblleft
pellicle mirror\textquotedblright\ is%
\begin{equation}
v=\frac{dx}{dt}=v_{p}\sin 2\omega _{p}t  \label{solution v(t)}
\end{equation}%
where the solution for the velocity amplitude is%
\begin{equation}
v_{p}=\omega _{p}x_{p}=\frac{1}{8}\frac{\varepsilon _{0}E_{p}^{2}}{m\omega
_{p}}  \label{v_p =...}
\end{equation}

Now from (\ref{Degenerate parametric oscillator threshold}), we found that
the threshold for the total mechanical energy of the driven
\textquotedblleft pellicle mirror\textquotedblright\ required to achieve
degenerate parametric oscillation is%
\begin{equation}
U_{\text{threshold}}=\frac{8m\omega _{s}^{2}d_{0}^{2}}{Q_{s}^{2}}
\label{threshold in total energy in Fig. 3}
\end{equation}%
where $m$ is the mass of the \textquotedblleft pellicle
mirror\textquotedblright , $\omega =\omega _{s}$ is the signal (or idler)
frequency, $d_{0}$ is the effective length in the $LC$ model of the signal
cavity, and $Q=Q_{s}$ is the quality factor of the signal (or idler) cavity
(i.e., the right cavity of Figure 3).

In steady state, and in the absence of any dissipation into heat, we infer
that the threshold total mechanical energy of the driven \textquotedblleft
pellicle mirror\textquotedblright\ will come into equilibrium with the
time-averaged threshold microwave energy stored in the left (pump) cavity of
Figure 3, i.e., 
\begin{equation}
U_{\text{threshold}}=\left\langle U_{p,\text{threshold}}\right\rangle
\label{equality of mechanical and EM energy in pump cavity}
\end{equation}%
Therefore, \ in order to convert (\ref{threshold in total energy in Fig. 3})
into a threshold pump power, we shall use the steady-state relationship (\ref%
{steady-state pump power}), which leads to the threshold condition \cite%
{consistency with (10) in FQMT11}%
\begin{equation}
\mathcal{P}_{\text{threshold}}=\frac{\omega _{p}U_{\text{threshold}}}{Q_{p}}=%
\frac{8m\omega _{p}\omega _{s}^{2}d_{0}^{2}}{Q_{s}^{2}Q_{p}}
\label{threshold power for no-DC-bias degenerate case}
\end{equation}%
This is the minimum amount of power needed to cause degenerate parametric
oscillations of signal and idler waves that start to build up exponentially
in the right cavity of Figure 3.

If we make the further simplifying assumption here that both the pump (left)
and signal (right) cavities of Figure 3 are similar superconducting
cavities, so that%
\begin{equation}
Q_{s}=Q_{p}=Q\sim 10^{10}
\end{equation}%
then (\ref{threshold power for no-DC-bias degenerate case}) simplifies to
the expression%
\begin{equation}
\mathcal{P}_{\text{threshold}}=\frac{8m\omega _{p}\omega _{s}^{2}d_{0}^{2}}{%
Q^{3}}\propto \frac{1}{Q^{3}}
\label{simplified threshold goes as inverse cube of Q}
\end{equation}%
i.e., the threshold pump power scales inversely as the \emph{cube} of the $Q$
of the superconducting cavities.

Let us now put in some numbers to see whether an experiment is feasible to
do or not. Suppose that $m=2$ milligrams (as in our FQMT11 paper \cite%
{consistency with (10) in FQMT11}), and that $\omega _{p}\simeq \omega
_{s}\simeq 2\pi \times 10^{10}$ Hz, and that $d_{0}\simeq \lambda _{s}$, so
that $\omega _{s}^{2}d_{0}^{2}=\omega _{s}^{2}\lambda _{s}^{2}\simeq 4\pi
^{2}c^{2}$. Then (\ref{simplified threshold goes as inverse cube of Q})
becomes%
\begin{eqnarray}
\mathcal{P}_{\text{threshold}} &\simeq &32\pi ^{2}\cdot mc^{2}\cdot \frac{%
\omega _{p}}{Q^{3}}  \notag \\
&\simeq &64\pi ^{3}\times 2\times 10^{-6}\text{kg}\times \left( 3\times
10^{8}\frac{\text{m}}{\text{s}}\right) ^{2}\times \frac{10^{10}\text{Hz}}{%
\left( 10^{10}\right) ^{3}}  \notag \\
&\simeq &3.6\text{ microwatts}  \label{3.6 microwatts}
\end{eqnarray}%
which indicates that the experiment sketched in Figure 3 is indeed feasible
to perform using our dilution refrigerators. The most critical and difficult
part of the experiment will be to ensure that the $Q$ of the superconducting
cavities will indeed be on the order of 10$^{10}$ or so.

Next, we shall consider the case in Figure 3 in which there exists a DC
bias, so that a DC electric field $E_{0}$ is then superposed with a
microwave pump field of amplitude $E_{p}$ of a pump wave at the second
harmonic of the signal. Therefore, let this pump have a waveform%
\begin{equation}
E=E_{0}-E_{p}\sin \omega _{p}t
\end{equation}%
where now the pump frequency $\omega _{p}$ is at the \emph{second} harmonic
(i.e., at \emph{twice} the frequency as the degenerate signal and idler
frequencies $\omega _{s}=\omega _{i}$). Then squaring this superposition of
fields to find the force in (\ref{Newton's 2nd law}), we obtain%
\begin{equation}
E^{2}=E_{0}^{2}-2E_{0}E_{p}\sin \omega _{p}t+E_{p}^{2}\sin ^{2}\omega _{p}t
\label{square of sum of E_0 and E_p}
\end{equation}%
We shall adopt\ throughout this calculation the $LC$ circuit model for the
pump cavity, as well as for the signal cavity, in which all electric fields,
like those in the capacitor $C$, are assumed to be uniform fields. If the DC
electric field is much larger than the amplitude of the microwave pump
waveform, i.e., if%
\begin{equation}
E_{0}>>E_{p}
\end{equation}%
then the dominant time-varying term of (\ref{square of sum of E_0 and E_p})
is the one at the first harmonic of $\omega _{p}$, so that (\ref{Newton's
2nd law}) becomes, to a good approximation,%
\begin{equation}
-\left( \frac{1}{2}\varepsilon _{0}\times 2E_{0}E_{p}\sin \omega
_{p}t\right) A=m\frac{d^{2}x}{dt^{2}}
\end{equation}%
Then the solution for the displacement $x$ of the plate is%
\begin{equation}
x=x_{p}\sin \omega _{p}t
\end{equation}%
with a displacement amplitude%
\begin{equation}
x_{p}=\frac{\varepsilon _{0}E_{0}E_{p}A}{m\omega _{p}^{2}}
\end{equation}%
Therefore the solution for the velocity $v$ of the plate is%
\begin{equation}
v=v_{p}\cos \omega _{p}t
\end{equation}%
with a velocity amplitude%
\begin{equation}
v_{p}=\frac{\varepsilon _{0}E_{0}E_{p}A}{m\omega _{p}}
\end{equation}%
Identifying this velocity amplitude with the velocity amplitude in (\ref%
{velocity at second harmonic is pump velocity amplitude}), we see that, for
this kind of \textquotedblleft external pumping agency,\textquotedblright\
i.e., the one using the pump in the left cavity of Figure 3 $with$ a DC bias,
the resulting driven velocity amplitude of the \textquotedblleft pellicle
mirror\textquotedblright\ is%
\begin{equation}
v_{2\omega }=v_{\text{pump}}=v_{p}=\frac{\varepsilon _{0}E_{0}E_{p}A}{%
m\omega _{p}}  \label{threshold pump velocity}
\end{equation}%
Solving for the pump electric field amplitude, one finds%
\begin{equation}
E_{p}=\frac{m\omega _{p}v_{p}}{\varepsilon _{0}E_{0}A}
\label{pump field amplitude in terms of v_p}
\end{equation}

To calculate the threshold power in the presence of a DC bias for parametric
oscillation of the degenerate signal and idler waves inside the right
cavity, we shall assume that the left cavity of Figure 3, which is filled
with pump waves, will fix the amplitude of the mechanical drive of the
\textquotedblleft pellicle mirror\textquotedblright , so that the pump
velocity amplitude (\ref{threshold pump velocity}) remains unchanged (i.e.,
undepleted) by the generation of weak signal and idler waves in the right
cavity. We shall then use the threshold condition (\ref{threshold in terms
of Q}) for the gain coefficient $\kappa $ that we obtained earlier for the
generation of degenerate signal and idler waves 
\begin{equation}
\kappa _{p,\text{threshold}}^{\text{(DC bias)}}=\frac{1}{4}\frac{v_{p,\text{%
threshold}}^{\text{(DC bias)}}}{d_{0}}=\frac{\omega _{s}}{Q_{s}}=\frac{1}{2}%
\frac{\omega _{p}}{Q_{s}}  \label{kappa threshold}
\end{equation}%
where $Q_{s}$ is the $Q$ factor of the right (signal) cavity of Figure 3.
Solving for the threshold pump velocity, one finds%
\begin{equation}
v_{p,\text{threshold}}^{\text{(DC bias)}}=\frac{2d_{0}\omega _{p}}{Q_{s}}
\label{threshold velocity}
\end{equation}%
Solving for the threshold electric field using (\ref{pump field amplitude in
terms of v_p}) and (\ref{threshold velocity}), we find%
\begin{equation}
E_{p,\text{threshold}}^{\text{(DC bias)}}=\frac{2m\omega _{p}^{2}d_{0}}{%
\varepsilon _{0}E_{0}AQ_{s}}  \label{threshold electric field}
\end{equation}%
Identifying the time-averaged threshold electric field energy density in the
pump cavity as%
\begin{equation}
\left\langle u_{E}\right\rangle _{p,\text{threshold}}^{\text{(DC bias)}}=%
\frac{1}{4}\varepsilon _{0}\left( E_{p,\text{threshold}}^{\text{(DC bias)}%
}\right) ^{2}
\end{equation}%
and expressing the total energy stored in the pump cavity, in the $LC$
capacitor model for the cavity, as%
\begin{eqnarray}
\left\langle U_{\text{tot}}\right\rangle _{p,\text{threshold}}^{\text{(DC
bias)}} &=&\left( \left\langle u_{E}\right\rangle _{p,\text{threshold}}^{%
\text{(DC bias)}}+\left\langle u_{B}\right\rangle _{p,\text{threshold}}^{%
\text{(DC bias)}}\right) Ad_{0}  \notag \\
&=&\frac{1}{2}\varepsilon _{0}\left( E_{p,\text{threshold}}^{\text{(DC bias)}%
}\right) ^{2}Ad_{0}
\end{eqnarray}%
and using the fact that this energy is related to the threshold input power
in steady state given by (\ref{steady-state pump power}), we conclude from (%
\ref{threshold electric field}) that the threshold pump power in the
presence of the DC bias is given by%
\begin{eqnarray}
\mathcal{P}_{\text{threshold}}^{\text{(DC bias)}} &=&\left\langle U_{\text{%
tot}}\right\rangle _{p,\text{threshold}}^{\text{(DC bias)}}\cdot \left( 
\frac{\omega _{p}}{Q_{p}}\right)  \notag \\
&=&\frac{1}{2}\left( \frac{1}{2}\varepsilon _{0}\left( E_{p,\text{threshold}%
}^{\text{(DC bias)}}\right) ^{2}\right) Ad_{0}\cdot \left( \frac{\omega _{p}%
}{Q_{p}}\right)  \notag \\
&=&\frac{1}{4}\varepsilon _{0}\left( \frac{2m\omega _{p}^{2}d_{0}}{%
\varepsilon _{0}E_{0}AQ_{s}}\right) ^{2}Ad_{0}\cdot \left( \frac{\omega _{p}%
}{Q_{p}}\right)  \notag \\
&=&\frac{m^{2}\omega _{p}^{5}d_{0}^{3}}{\varepsilon
_{0}E_{0}^{2}AQ_{s}^{2}Q_{p}}  \label{power threshold in terms of E_0}
\end{eqnarray}%
If we again make the simplifying assumption that%
\begin{equation}
Q_{p}=Q_{s}=Q
\end{equation}%
because both the left and right cavities are superconducting cavities with
similar configurations, then (\ref{power threshold in terms of E_0}) becomes%
\begin{equation}
\mathcal{P}_{\text{threshold}}^{\text{(DC bias)}}=\frac{m^{2}\omega
_{p}^{5}d_{0}^{3}}{\varepsilon _{0}E_{0}^{2}AQ^{3}}\propto \frac{1}{Q^{3}}
\label{power threshold for identical Q's}
\end{equation}%
i.e., the threshold once again depends on the inverse \emph{cube} of the
quality factor of the superconducting cavities. Since for the TM 011 mode,
the distance scale for the capacitor gap is approximately%
\begin{equation}
d_{0}\simeq \frac{\lambda _{p}}{2}=\frac{\pi c}{\omega _{p}}
\end{equation}%
the factor $\omega _{p}^{5}d_{0}^{3}$ simplifies to become%
\begin{equation}
\omega _{p}^{5}d_{0}^{3}\simeq \omega _{p}^{2}\pi ^{3}c^{3}
\end{equation}%
Hence the threshold pump power simplifies to become%
\begin{equation}
\mathcal{P}_{\text{threshold}}^{\text{(DC bias)}}\simeq \pi ^{3}\frac{%
m^{2}\omega _{p}^{2}c^{3}}{\varepsilon _{0}E_{0}^{2}AQ^{3}}=\pi ^{3}\frac{%
m^{2}c^{4}\omega _{p}^{2}}{\varepsilon _{0}E_{0}^{2}AcQ^{3}}
\label{threshold with DC bias and equal Q's}
\end{equation}%
If we now take the ratio of the expression with the DC bias (\ref{threshold
with DC bias and equal Q's}) with the expression for the case without the DC
bias (\ref{3.6 microwatts}), we see that%
\begin{eqnarray}
\frac{\mathcal{P}_{\text{threshold}}^{\text{(DC bias)}}}{\mathcal{P}_{\text{%
threshold}}^{{}}} &\simeq &\left( \pi ^{3}\frac{m^{2}c^{4}\omega _{p}^{2}}{%
\varepsilon _{0}E_{0}^{2}AcQ^{3}}\right) \cdot \left( 32\pi ^{2}\cdot
mc^{2}\cdot \frac{\omega _{p}}{Q^{3}}\right) ^{-1}  \notag \\
&=&\frac{\pi }{32}\frac{mc^{2}}{\varepsilon _{0}E_{0}^{2}A(c/\omega _{p})}=%
\frac{\pi ^{2}}{32}\frac{mc^{2}}{\varepsilon _{0}E_{0}^{2}Ad_{0}}  \notag \\
&=&\frac{\pi ^{2}}{32}\frac{mc^{2}}{\varepsilon _{0}E_{0}^{2}V_{0}}
\label{ratio of powers}
\end{eqnarray}%
where $V_{0}=Ad_{0}$ is the volume of the pump cavity. It is obviously the
case that the rest mass of the \textquotedblleft pellicle
mirror\textquotedblright\ is much larger by many orders of magnitude than
the energy stored in the DC electric field inside the pump cavity, i.e.,%
\begin{equation}
mc^{2}>>\varepsilon _{0}E_{0}^{2}V_{0}
\end{equation}%
Hence we definitely should \emph{not} use the DC bias method for trying to
achieve the threshold of the degenerate parametric oscillator. Nevertheless,
the use of an adjustable DC bias as indicated in Figure 3 could be useful as
an adjustable experimental knob for tuning the two microwave cavities into
resonance with respect to each other.\\

{\bf Acknowledgments:} I'd like to thank Jay Sharping, Luis Martinez and Robert Haun for their help in some preliminary room-temperature experiments to test the idea of microwave-driven pellicle-mirror motion arising from the Maxwell stress tensor, and Lin Tian for helpful comments.


\begin{thebibliography}{99}
\bibitem{alternative} One can derive (\ref{pressure on plate}) in the
following way: An instantaneous force $dF$ on a given capacitor plate arises
from adding an instantaneous charge $dQ$ to this plate in the presence of
the instantaneous electric field $E$ inside the capacitor through the
relationship%
\begin{equation}
dF=dQ\cdot E  \label{force = charge x E field}
\end{equation}%
However, by Gauss's law applied to a pillbox staddling the inner surface of
the fixed left plate, one finds that the instantaneous electric field in
between the capacitor plates is given by%
\begin{equation}
E=\frac{\sigma }{\varepsilon _{0}}=\frac{Q}{\varepsilon _{0}A}
\label{E = Q/A}
\end{equation}%
where $\sigma =Q/A$ is the charge density on the plate of area $A$. From (%
\ref{E = Q/A}), it follows that%
\begin{equation}
dQ=dE\cdot \left( \varepsilon _{0}A\right)
\end{equation}%
Therefore the incremental force $dF$ on the plate upon the increment of the
charge $dQ$\ becomes%
\begin{equation}
dF=dQ\cdot E=E\cdot dE\cdot \left( \varepsilon _{0}A\right)
\end{equation}%
By integration, one finds that the total force is given by%
\begin{equation}
F=\left( \varepsilon _{0}A\right) \int\limits_{0}^{E}EdE=\left( \frac{1}{2}%
\varepsilon _{0}E^{2}\right) A
\end{equation}%
and therefore that the pressure (i.e., a component of the Maxwell stress
tensor) is given by%
\begin{equation}
P=\frac{F}{A}=\frac{1}{2}\varepsilon _{0}E^{2}=u_{E}
\end{equation}%
where $u_{E}$ is the instantaneous energy density stored in the electric
field inside the capacitor. From this derivation, it is clear that it is the
$Coulomb$ force, and not the $Lorentz$ $\mathbf{v\times B}$ force associated
with radiation pressure, which is operative in driving the motion of the
moving mirror in the TM mode pictured in Figure 1. This implies that the
Maxwell stress tensor (\ref{Maxwell stress tensor}) should be much more
effective than radiation pressure in producing parametric gain and
oscillation of a moving mirror. Now consider the lowest possible transverse
magnetic mode, i.e., the TM 010 mode, which is a mode with longitudinal
electric field lines that terminate on the charges at the end faces of the
cavity, as indicated in Figure 1. These charges are necessary for there to
exist a longitudinal force associated with the Maxwell stress tensor, which
acts along the axis of the pellicle mirror to drive it into motion. By
contrast, the lowest possible transverse electric mode, i.e., the TE 111
mode, has transverse, but no longitudinal, electric field lines. Hence no
longitudinal force can arise from the charges induced by this TE mode.
However, there does exist a longitudinal radiation pressure force arising
from the Poynting vector at the surface of the \textquotedblleft pellicle
mirror\textquotedblright\ for the TE modes, but it is much smaller in
magnitude than the longitudinal force arising from the Maxwell stress tensor
for the TM modes.
\bibitem{sign of work} The sign of the work as defined here in (\ref{work
done on moving capacitor plate}) and (\ref{PdV work done by piston}) is
opposite to the sign of the work as defined in thermodynamics, which is the
work done \emph{by} a piston \emph{on} an external agency. In
thermodynamics, by convention, the sign of the work is \emph{positive} when
the work is done \emph{by} the \textquotedblleft system\textquotedblright\ 
\emph{on} the \textquotedblleft environment,\textquotedblright\ i.e., when
useful thermodynamic work can be extracted \emph{from} the \textquotedblleft
system\textquotedblright\ \emph{via} the piston \emph{into} the
\textquotedblleft environment.\textquotedblright\ Here, by contrast, the
sign of the mechanical work in (\ref{work done on moving capacitor plate})
and (\ref{PdV work done by piston}) is \emph{positive} when the work is done 
\emph{by} the external agency (i.e., the pump) \emph{on} the capacitor plate
(i.e., the fields of the $LC$ circuit), so that useful parametric \emph{%
amplification} can occur.
\end{thebibliography}
\end{document}